\documentstyle[12pt]{article}

\def\a {{\alpha}}

\def\a {{\alpha}}

\def\bn {\begin{eqnarray}}
\def\en {\end{eqnarray}}
\newcommand{\be}{\begin{equation}}
\newcommand{\ee}{\end{equation}}
\newcommand{\ba}{\begin{eqnarray}}
\newcommand{\ea}{\end{eqnarray}}

\begin{document}
\begin{center}
 {\bf\Large{On fractional Euler-Lagrange and Hamilton equations
 and the fractional generalization of total time derivative
  }}
\end{center}
\begin{center}
{\bf Dumitru Baleanu}\footnote[1]{On leave of absence from
Institute of Space Sciences, P.O.BOX, MG-23, R 76900,
Magurele-Bucharest, Romania,E-mails: dumitru@cankaya.edu.tr,
baleanu@venus.nipne.ro}

Department of Mathematics and Computer Sciences, Faculty of Arts
and Sciences, \c{C}ankaya University- 06530, Ankara, Turkey\\

 {\bf
Sami I. Muslih}\footnote[2]{E-mail: smuslih@ictp.trieste.it}\\
 Department of Physics, Al-Azhar
University, Gaza, Palestine\\
and\\
 International Center for Theoretical Physics,Trieste,
Italy

 {\bf Eqab M. Rabei}\\
Department of Science-Jerash Private University,
Jerash-Jordan\\and\\ Department of Physics-Mutah University,
Karak-Jordan

\end{center}

\begin{abstract}
Fractional mechanics describes both conservative and
non-conservative systems. The fractional variational principles
gained importance in studying the fractional mechanics and several
versions are proposed. In classical mechanics the equivalent
Lagrangians play an important role because they admit the same
Euler-Lagrange equations. By adding a total time derivative of a
suitable function to a given classical Lagrangian or by
multiplying with a constant the Lagrangian we obtain the same
equations of motion.
 In this study, the fractional discrete  Lagrangians which differs by a fractional derivative  are analyzed within
Riemann-Liouville fractional derivatives.
 As a consequence of applying this procedure the classical results
 are reobtained as a special case.

The fractional generalization of $Fa\grave{a}$ di Bruno formula is
used in order to obtain the concrete expression of the fractional
 Lagrangians which differs from a given fractional Lagrangian by adding a fractional derivative.
  The fractional Euler-Lagrange and Hamilton
equations corresponding to the obtained fractional Lagrangians are
investigated and two examples are analyzed in details.

\end{abstract}

\section{Introduction}

Fractional calculus is a generalization of ordinary
differentiation and integration to arbitrary order.The fractional
derivatives are the infinitesimal generators of a class of
translation invariant convolution semigroups which appear
universally as attractors. Various applications of fractional
calculus are based on replacing the time derivative in an
evolution equation with a derivative of fractional order. The
results of several recent researchers confirm that fractional
derivatives seem to arise for important mathematical reasons.
During the last decades the fractional calculus [1-3] started to
be used in various fields, e.g. engineering, physics, biology and
many important results were reported [4-19]. A special attention
has been devoted during the last years to the fractional
variational principles [20-29] and their applications [30-35]. For
the constrained systems \cite{henneaux} the fractional Lagrangian
and Hamiltonian formalism are still at the beginning of their
development.
 The
formulation of the fractional variational principles still needs
to be more elaborated in the future and it will have an important
impact on the elaboration of a consistent  fractional quantization
method for both discrete and continuous systems.The fractional
variational principles  is deeply related to the fractional
quantization procedure. For a given Lagrangian there are several
proposed methods to obtain the fractional Euler-Lagrange equations
and the corresponding Hamiltonians. However this issue is not yet
complectly clarified and it requires more further detailed
analysis. One of the main obstacles is related to the fractional
Leibniz rule, the fractional chain rule \cite{abramoviz} as well
as the fractional Taylor series \cite{hardy,truijilo}. The second
problem is related to the non-locality of the fractional
Lagrangian.Recently, the Hamiltonian formalism for nonlocal
Lagrangians was investigated in \cite{llosa} and a relation of
non-local theories and Ostrogradski's formalism \cite{gitman,
nesterenko} was reported \cite{bering},\cite{jgomis}.

As it is well known the classical equivalent Lagrangians represent
a powerful tool for studying the Hamilton-Jacobi equation in
$Carath\acute{e}odory's$ formulation \cite{cara} as well as in
generalized mechanics \cite{negri,santili}.

For these reasons, in this study we generalized the notion of
equivalent Lagrangian for the fractional case. The fractional
Euler-Lagrange equations of fractional equivalent Lagrangians are
calculated and the fractional Hamiltonians are constructed.

The plan of this paper is as follows:

In Section 2, some basic formulas of the fractional calculus are
briefly reviewed. Section 3 briefly review the fractional
Hamiltonian approach of discrete systems. In Section 4 the
fractional Lagrangians are investigated and two examples are
investigated. Section 5 is dedicated to our conclusions.

\section{Brief overview of fractional calculus}

 In this section, we formulate the problem in terms
of the left and the right Riemann-Liouville (RL) fractional
derivatives, which are defined as follows:  the~ left~
Riemann~-Liouville~ fractional~ derivative

\begin{equation}
{{}_a\textbf{D}_t^{\alpha}f(t)} =
\frac{1}{\Gamma{(n-\alpha)}}\left(\frac{d}{dt}\right)^{n}\int\limits_a^t(-\tau+t)^{n-\alpha-1}f(\tau)d\tau,
\end{equation}

and the~ right ~Riemann~-Liouville ~fractional~ derivative

\begin{equation}
{{}_t\textbf{D}_b^{\alpha}f(t)} =
\frac{1}{\Gamma{(n-\alpha)}}\left(-\frac{d}{dt}\right)^{n}\int\limits_t^b(\tau-t)^{n-\alpha-1}f(\tau)d\tau,
\end{equation}
where the order $\alpha$ fulfills $n-1\leq\alpha <n$ and $\Gamma$
represents the Euler's  Gamma function. It is observed that if
$\alpha$ becomes an integer, we recovered  the usual definitions,
namely,
\begin{equation}
{{}_a\textbf{D}_t^{\alpha}}f(t)
=\left(\frac{d}{dt}\right)^{\alpha}f(t),~~{{}_t\textbf{D}_b^{\alpha}}f(t)
= \left(-\frac{d}{dt}\right)^{\alpha}f(t), ~\a=1,2,... .
\end{equation}
Fractional RL derivatives have many interesting properties. By
direct calculation we observe that the RL derivative of a constant
is not zero, namely
\begin{equation}\label{ce}
{}_a\textbf{D}_t^{\alpha}C=C\frac{(t-a)^{-\alpha}}{\Gamma(1-\alpha)}.
\end{equation}
RL derivative of a power of t has the following form
\begin{equation}
{}_a\textbf{D}_t^{\alpha}t^\beta=\frac{\Gamma(\alpha+1)t^{\beta-\alpha}}{\Gamma(\beta-\alpha+1)},
\end{equation}
for $\alpha> -1, \beta\geq 0$. Composite of fractional derivatives
is given by the following formula

\begin{equation}\label{com}
{}_a\textbf{D}_t^{\alpha}{}_a\textbf{D}_t^{\sigma}f(t)={}_a\textbf{D}_t^{\alpha+\sigma}f(t)-\sum_{j=1}^{k}{}_a\textbf{D}_t^{\sigma-j}f(t)|_{t=a}
\frac{(t-a)^{-\alpha-j}}{\Gamma(1-\alpha-j)},
\end{equation}
where $0\leq k-1\leq q\leq k$, $p\geq 0$ and k is a whole number.
As it can be seen from (\ref{com}) the composition is not
commutative. Finally,  the fractional product rule is given below
\begin{equation}\label{com1}
{}_a\textbf{D}_t^{\alpha}(fg)=\sum_{j=0}^{\infty}\left(%
\begin{array}{c}
  \alpha \\
  j \\
\end{array}%
\right)\left({}_a\textbf{D}_t^{\alpha-j}f\right)\left(\frac{d^i
g}{d t^j} \right).
\end{equation}

By inspection  we observe that the fractional product contains
infinitely many terms and this product is taking into account the
memory.

\subsection{Fractional derivative of a composite function}

In order to find the fractional generalization of a composition
function we have obtain first of all the most general classical
counterpart.

  Let us take an analytic function $\phi(t)$ and
$f(t)=H(t-a)$, where $H(t)$ is the Heaviside function. Using the
Leibniz rule and the formula for the fractional differentiation of
the Heaviside function we obtain
\begin{equation}
{}_a\textbf{D}_t^{p}\phi(t)=\sum_{k=0}^{\infty}\left(
                                                 \begin{array}{c}
                                                   p \\
                                                   k \\
                                                 \end{array}
                                               \right)
                                               \phi^{(k)}(t){}_a\textbf{D}_t^{p-k}H(t-a)
\end{equation}
or

\begin{equation}\label{im}
{}_a\textbf{D}_t^{p}\phi(t)=\frac{(t-a)^{-p}}{\Gamma(1-p)}\phi(t)
+\sum_{k=1}^{\infty}\left(
                      \begin{array}{c}
                        p \\
                        k \\
                      \end{array}
                    \right)
                    \frac{(t-a)^{k-p}}{\Gamma(k-p+1)}\phi^{(k)}(t)
\end{equation}under the assumption $t > a$.
Let us suppose that $\phi(t)$ is a composite function

\begin{equation}
\phi(t)=F(h(t)).
\end{equation}

The k-th order derivative of $\phi(t)$ is evaluated with the help
of the $Fa\grave{a}$ di Bruno formula \cite{abramoviz}

\begin{equation}
\frac{d^k}{dt^k}F(h(t))=k!\sum_{m=1}^k
F^{(m)}(h(t))\sum\prod_{r=1}^{k}\frac{1}{a_r!}\left(\frac{h^{(r)}(t)}{r!}\right)^{a_r},
\end{equation}
where the sum $\sum$ extends over all combinations of non-negative
integer values of $a_1,\cdots ,a_k$ such that
\begin{equation}
\sum_{r=1}^{k}ra_r=k
\end{equation}
and
\begin{equation}
\sum_{r=1}^{k}a_r=m.
\end{equation}
Therefore, the fractional derivative of a composition function is
given by
\begin{eqnarray}\label{diBruno}
&{}_a\textbf{D}_t^{p}F(h(t))&=\frac{(t-a)^{-p}}{\Gamma(1-p)}F(h(t))+\cr
&\sum_{k=1}^{\infty}\left(
                      \begin{array}{c}
                        p \\
                        k \\
                      \end{array}
                    \right)
&k!\frac{(t-a)^{k-p}}{\Gamma(k-p+1)}\sum_{m=1}^{k}F^{(m)}(h(t))\sum\prod_{r=1}^{k}\frac{1}{a_r!}\left(\frac{h^{(r)}(t)}{r!}\right)^{a_r},
\end{eqnarray}
where the sum $\sum$ and coefficients $a_r$ have the meaning
explained above \cite{podlubny}.

\section{Fractional Lagrangian and Hamiltonian \\analysis of discrete systems}

 For a given fractional Lagrangian given by

\begin{equation}
L_f(q^{\rho}(t),{}_a\textbf{D}_t^{\alpha}q^{\rho}(t),
{}_t\textbf{D}_b^{\alpha}q^\rho(t)),~\rho=1,\cdots,N,
\end{equation}

the Euler-Lagrange equations are given as follows (see for example
Refs. \cite{8},\cite{balsami1} and the references therein)

\begin{eqnarray}\label{eeeee}
\frac{\partial L_f}{\partial q^{\rho}(t)}+
{}_t\textbf{D}_b^{\alpha}\frac{\partial L_f}{\partial
{}_a\textbf{D}_t^{\alpha}q^{\rho}(t) } +
{}_a\textbf{D}_t^{\alpha}\frac{\partial L_f}{\partial
{}_t\textbf{D}_b^{\alpha}q^{\rho}(t) }=0,~0 < \alpha< 1.
\end{eqnarray}

For simplicity,in the following we consider the following form of
the fractional Euler-Lagrange equations

\begin{eqnarray}\label{eeeeee}
\frac{\partial L^{'}_f}{\partial q^{\rho}(t)}+
{}_t\textbf{D}_b^{\alpha}\frac{\partial L^{'}_f}{\partial
{}_a\textbf{D}_t^{\alpha}q^{\rho}(t)}=0,~0 < \alpha<
1~,\rho=1,\cdots, N.\end{eqnarray}

In the following by using (\ref{eeeeee}) we define the generalized
momenta as (see Ref.\cite{eqab} for more details)

\begin{equation}\label{fd1}
p_{\alpha_\rho}=\frac{\partial L_f^{'}}{\partial
{}_a\textbf{D}_t^{\alpha}q^{\rho}(t) }, \rho=1,\cdots,N.
\end{equation}
As a consequence of (\ref{eeeeee}) and (\ref{fd1}) a Hamiltonian
function is defined as
\begin{equation}\label{fd3}
H=p_{\alpha_\rho}{}_a\textbf{D}_t^{\alpha}q^{\rho}(t) - L_f^{'}.
\end{equation}
The canonical equations corresponding to (\ref{fd3}) are given below

\begin{equation}\label{fda33}
\frac{\partial H}{\partial t}=-\frac{\partial L_f^{'}}{\partial
t},~ \frac{\partial H}{\partial
p_{\alpha_\rho}}={}_a\textbf{D}_t^{\alpha}q^{\rho},~\frac{\partial
H}{\partial q^{\rho}}={}_t\textbf{D}_b^{\alpha}p_{\alpha_\rho}, 0<
\alpha < 1,\rho=1,\cdots,N.
\end{equation}
We mention that other interesting formulations of fractional
Lagrangian and Hamiltonian dynamics can be found in \cite{4,5} and
\cite{klimek1}.

\section{Fractional derivatives and fractional Lagrangian mechanics}
Let $L(q^{\rho}(t),{\dot q^{\rho}}(t))$ be a classical Lagrangian
function  with $\rho=1,\cdots,N$. Let $L^{'}(q^{\rho},{\dot
q^{\rho}(t)})=L(q^{\rho}(t),{\dot
q^{\rho}}(t))+\frac{dF(q^{m}(t))}{dt}$, where $F(q^{m}(t))$ is a
differentiable function and $q^{m}(t)$ is one of the coordinates.
It is very well known that two equivalent classical Lagrangians
admit the same Euler-Lagrange equations. In the following we are
considering the fractional generalization of this classical
results, namely we replace the normal derivatives by the
fractional ones.
\subsection{A particular case}
One of the possible generalizations of the classical Lagrangian
$L(q^{\rho}(t),{\dot q^{\rho}}(t))$ is given by
\begin{equation}\label{ffrac}
L_f(q^{\rho}(t),{}_a\textbf{D}_t^{\alpha}q^{\rho}(t)).
\end{equation}
If we add to (\ref{ffrac}) a term of the form
${}_t\textbf{D}_b^{\alpha}q(t)$ we obtain
\begin{equation}\label{ecuf}
L_f^{'}=L_f(q^{\rho}(t),{}_a\textbf{D}_t^{\alpha}q^{\rho}(t))+
C{}_a\textbf{D}_t^{\alpha}q^{m}(t),
\end{equation}

where $m\in\{1,\cdots,N\}$ is a chosen coordinate and C is a
non-zero real constant.

The fractional Euler-Lagrange equations of (\ref{ecuf}) are given
by
\begin{equation}
\frac{\partial L_f}{\partial q^{\sigma}(t)}+
{}_t\textbf{D}_b^{\alpha}\frac{\partial L_f}{\partial
{}_a\textbf{D}_t^{\alpha}q^{\sigma}(t) }=0, \end{equation}
\begin{equation}\label{ecufin}
\frac{\partial L_f}{\partial q^{m}(t)}+
{}_t\textbf{D}_b^{\alpha}\frac{\partial L_f}{\partial
{}_a\textbf{D}_t^{\alpha}q^{m}(t) }+
C\frac{(b-t)^{-\alpha}}{\Gamma(1-\alpha)}=0, \end{equation}
\begin{equation}
\frac{\partial L_f}{\partial q^{\delta}(t)}+
{}_t\textbf{D}_b^{\alpha}\frac{\partial L_f}{\partial
{}_a\textbf{D}_t^{\alpha}q^{\delta}(t) }=0,
\end{equation}

where $\sigma=1,\cdots, m-1$ and $\delta=m+1,\cdots, N$.

The last term of (\ref{ecufin}) arises after taking into account
(\ref{eeeee}) and (\ref{ce}) with $C=1$.

\subsubsection{An example} Let us consider the fractional generalization of  a free
Lagrangian of one degree of freedom $L=\frac{{\dot x}^2}{2}$ at
which we add a term of the form ${\dot x}$. At the classical level
the Lagrangians $L=\frac{{\dot x}^2}{2}$ and $L^{'}=\frac{{\dot
x}^2}{2}+ {\dot x}$ are equivalent.

One of the possible generalizations for the fractional case is
given below
\begin{equation}\label{er}
L_f=\frac{({}_a\textbf{D}_t^{\alpha}q(t))^2}{2}
+C{}_a\textbf{D}_t^{\alpha}q(t),
\end{equation}
where C is a non-zero real constant.

By using (\ref{er}) the fractional equation of motion is given
bellow

\begin{equation}\label{ecufin1}
 {}_t\textbf{D}_b^{\alpha}
({}_a\textbf{D}_t^{\alpha}q(t)) +
C\frac{(b-t)^{-\alpha}}{\Gamma(1-\alpha)}=0. \end{equation}

From (\ref{er}) the expression of the fractional canonical
momentum is

\begin{equation}\label{edf}
p_\alpha= {}_a\textbf{D}_t^{\alpha}q(t) + C.
\end{equation}

The fractional canonical Hamiltonian is given by
\begin{equation}\label{hf}
H_f=p_\alpha
{}_a\textbf{D}_t^{\alpha}q(t)-L_f=\frac{(p_\alpha-C)^2}{2}.
\end{equation}

Applying the method developed in \cite{eqab} we obtain
\begin{equation}\label{ecufin2}
\frac{\partial H}{\partial p_\alpha}=
{}_a\textbf{D}_t^{\alpha}q(t), \frac{\partial H}{\partial q}=
{}_t\textbf{D}_b^{\alpha}p_\alpha ,
\end{equation}
or
\begin{equation}\label{ecufin21}
{}_a\textbf{D}_t^{\alpha}q(t)=p_\alpha -
C,~{}_t\textbf{D}_b^{\alpha}p_\alpha=0.
\end{equation}

Therefore, from (\ref{ecufin21}) and (\ref{ecufin1}) and taking
into account (\ref{edf}) we conclude that fractional
Euler-Lagrange and fractional Hamilton equations give the same
evolution equation for the coordinate q(t). Besides, when the
constant C becomes zero and $\alpha\rightarrow 1$ the classical
results are reobtained.

\subsection{The general case}
 The next step is to add to the  fractional Lagrangian
$L_f(q^{\rho}(t),{}_a\textbf{D}_t^{\alpha}q^{\rho}(t)$),
$\rho=1,\cdots,N$ a most general term which under some limits
becomes the classical total derivative. We analyze the fractional
equivalent Lagrangians as it is given below
\begin{equation}\label{fractot}
L_f^{'}=L_f(q^{\rho}(t),{}_a\textbf{D}_t^{\alpha}q^{\rho}(t))+
{}_a\textbf{D}_t^{\alpha}F(q^{m}(t)),~\rho=1,\cdots,n.
\end{equation} By using the fractional generalization of
$Fa\grave{a}$ di Bruno formula given in (\ref{diBruno}) we obtain
the explicit form of (\ref{fractot}) as follows

\begin{eqnarray}\label{ffrac1}
&L_f^{'}=&L_f(q^{\rho}(t),{}_a\textbf{D}_t^{\alpha}q^{\rho}(t))+
\frac{(t-a)^{-\alpha}}{\Gamma(1-\alpha)}F(q^{m}(t))+\cr
&\sum_{k=1}^{\infty}\left(
                      \begin{array}{c}
                        \alpha \\
                        k \\
                      \end{array}
                    \right)
&k!\frac{(t-a)^{k-\alpha}}{\Gamma(k-\alpha+1)}\sum_{s=1}^{k}F^{(s)}(q^{m}(t))\sum\prod_{r=1}^{k}\frac{1}{a_r!}\left(\frac{(q^{m})^{(r)}(t)}{r!}\right)^{a_r}.
\end{eqnarray}

The presence of the infinite higher order derivatives in
(\ref{ffrac1}) rises an interesting question regarding the form of
the  corresponding Euler-Lagrange equations as well as the form of
the corresponding Hamiltonian construction. The theory described
by (\ref{ffrac1}) is non-local and involve infinity derivative of
$q(t)$. The key point is to use the formula (\ref{im}) which
allowed to represent the both parts of the Lagrangian in a common
way.

The next step is to write the Euler-Lagrange equations and the
corresponding Hamiltonian.

 In this line of taught
 we consider that the dynamical
variable $q(t)$ is a 1+1 dimensional field $Q(x,t)$ subjected to
the following chirality condition \cite{bering}

\begin{equation}\label{chiral}
\frac{dQ(x,t)}{dt}=\partial_xQ(x,t).
\end{equation}
A specific feature of the Hamiltonian formalism for non-local
theories is that is contains the Euler-Lagrange equations as
Hamiltonian constraints.

 By using (\ref{chiral}) we have
\begin{equation}
(\frac{d}{dt})^nQ(x,t)=(\partial_x)^nQ(x,t), n\in N_0,
\end{equation}
and having in mind that $Q(x,t)=q(x+t)$ assures the one-to-one
correspondence between $q(t)$ and $Q(x,t)$ \cite{bering}.
 Ostrogradski's
coordinates are defined as follows:

\begin{equation}
Q^{(n)}(t)=(\partial_x)^nQ(x,t)\mid_{x=x_0},
\end{equation}
where the discontinuity curve $x_0(t)=x_0$ is a constant
\cite{bering}.

 By using the inverse relation provided by the Taylor
expansion around $x=x_0$ we obtain

\begin{equation}
Q(x,t)=\sum_{n=0}^{\infty}\frac{(x-x_0)^n}{n!}Q^{(n)}(t).
\end{equation}
We notice that we reobtain the chirality condition as
\begin{equation}
\dot{Q}^{(n)}(t)=Q^{(n+1)}(t).
\end{equation}

A boundary Poisson bracket was introduced  in \cite{bering} and it
has the expression as follows

\begin{equation}\label{pois}
\{F(t),G(t)\}=\sum_{k,l=0}^{\infty}c_{k,l}\int_{-\infty}^{\infty}dx(\partial_x)^{k+l}[\frac{\delta
F(t)}{\delta Q^{(k)}}(x,t)\frac{\delta G(t)}{\delta
P^{(l)}}(x,t)]-(F\leftrightarrow G),
\end{equation}
where the coefficients $c_{k,l}$ are constants and normalized in
such a manner to satisfy the Jacobi identity. Fixing the time t
the classical canonical relation is naturally obtained from
(\ref{pois}) as
\begin{equation}\label{can}
\{Q(x,t),P(x^{'},t)\}=\delta_R(x-x^{'}).
\end{equation}

By using (\ref{pois}) and (\ref{can}) and defining Ostrogradski's
momenta $P_{(n)}(t)$ as
\begin{equation}\label{pe1}
P_{(n)}(t)=\int_{-\infty}^{\infty}dx\frac{(x-x_0)^n}{n!}P(x,t).
\end{equation}

From (\ref{pe1}) the form of $P(x,t)$ is given by
\begin{equation}\label{pe}
P(x,t)=\sum_{n=0}^{\infty}P_{(n)}(t)(-\partial_x)^n\delta_R(x-x_0).
\end{equation}

By using (\ref{pe}) the expression for $P_{(n)}(t)$ is as follows
\begin{equation}
P_{(n)}(t)=\sum_{m=n}^{\infty}(-\partial_t)^{m-n}\frac{\partial
L_f^{'}[Q](t)}{\partial Q^{(m+1)}(t)}.
\end{equation}

The above definitions leads us to the following Hamilton equations
\begin{equation}
{\dot P_{(n)}}(t)+ P_{(n-1)}(t)=\frac{\partial
L_f^{'}[Q](t)}{\partial Q^{(n)}(t)}, n\in N,
\end{equation}
\begin{equation}\label{el}
{\dot P_{(0)}}(t)=\frac{\partial L_f^{'}[Q](t)}{\partial
Q^{(0)}(t)}.
\end{equation}
Taking into account the definition of ${ P_{(0)}}(t)$  we observe
that  (\ref{el}) is the Euler-Lagrange equation corresponding to
the lagrangian $L_f^{'}[Q](t)$ written in the compact form.
Finally, the Hamiltonian corresponding to $L_f^{'}$ is given below
\begin{equation}
H=\sum_{n=0}^{\infty} P_{(n)}(t)Q^{(n+1)}(t)-L_f^{'}[Q](t).
\end{equation}

Let us consider the following particular case

\begin{equation}\label{hor}
L_f^{'}=\frac{1}{2}({}_a\textbf{D}_t^{\alpha}q(t))^{2}+
{}_a\textbf{D}_t^{\alpha}F(q(t)).
\end{equation}
By using (\ref{im}) we obtain

\begin{eqnarray}
L_f^{'}&=&\frac{1}{2}\left(\frac{(t-a)^{-\alpha}}{\Gamma(1-\alpha)}q(t)
+\sum_{k=1}^{\infty}\left(
                      \begin{array}{c}
                        \alpha \\
                        k \\
                      \end{array}
                    \right)
                    \frac{(t-a)^{k-\alpha}}{\Gamma(k-\alpha+1)}q^{(k)}(t)\right)^2
                    +\frac{(t-a)^{-\alpha}}{\Gamma(1-\alpha)}F(q(t))\cr+
&\sum_{k=1}^{\infty}\left(
                      \begin{array}{c}
                        \alpha \\
                        k \\
                      \end{array}
                    \right)
&k!\frac{(t-a)^{k-\alpha}}{\Gamma(k-\alpha+1)}\sum_{s=1}^{k}F^{(s)}(q(t))\sum\prod_{r=1}^{k}\frac{1}{a_r!}\left(\frac{q^{(r)}(t)}{r!}\right)^{a_r}.
\end{eqnarray}
By making in (\ref{hor}) the corresponding substitutions of q(t)
into $Q(x,t)$ we obtain

\begin{eqnarray}\label{final}
L_f^{'}(Q)[t]&=&\frac{1}{2}\left(\frac{(t-a)^{-\alpha}}{\Gamma(1-\alpha)}Q(t,x)
+\sum_{k=1}^{\infty}\left(
                      \begin{array}{c}
                        \alpha \\
                        k \\
                      \end{array}
                    \right)
                    \frac{(t-a)^{k-\alpha}}{\Gamma(k-\alpha+1)}Q^{(k)}(t,x)\right)^2
                    + \frac{(t-a)^{-\alpha}}{\Gamma(1-\alpha)}F(Q(t,x))\cr
                    &+&
\sum_{k=1}^{\infty}\left(
                      \begin{array}{c}
                        \alpha \\
                        k \\
                      \end{array}
                    \right)
k!\frac{(t-a)^{k-\alpha}}{\Gamma(k-\alpha+1)}\sum_{s=1}^{k}F^{(s)}(Q(t,x))\sum\prod_{r=1}^{k}\frac{1}{a_r!}\left(\frac{Q^{(r)}(t,x)}{r!}\right)^{a_r}.
\end{eqnarray}

By using (\ref{final}) the corresponding  fractional
Euler-Lagrange equation has the form

\begin{eqnarray}\label{compactform}
 {\dot
P_{(0)}}&=&\frac{(t-a)^{-\alpha}}{\Gamma(1-\alpha)}\left(\frac{(t-a)^{-\alpha}}{\Gamma(1-\alpha)}Q(t,x)
+\sum_{k=1}^{\infty}\left(
                      \begin{array}{c}
                        \alpha \\
                        k \\
                      \end{array}
                    \right)
                    \frac{(t-a)^{k-\alpha}}{\Gamma(k-\alpha+1)}Q^{(k)}(t,x)\right)\cr
                    &+&\frac{(t-a)^{-\alpha}}{\Gamma(1-\alpha)}\frac{dF(Q(t,x))}{dQ(t,x)}\cr
&+& \sum_{k=1}^{\infty}\left(
                      \begin{array}{c}
                        \alpha \\
                        k \\
                      \end{array}
                    \right)
k!\frac{(t-a)^{k-\alpha}}{\Gamma(k-\alpha+1)}\sum_{s=1}^{k}\frac{dF^{(s)}(Q(t,x))}{dQ(t,x)}\sum\prod_{r=1}^{k}\frac{1}{a_r!}\left(\frac{Q^{(r)}(t,x)}{r!}\right)^{a_r}.
\end{eqnarray}
In the compact form (\ref{compactform}) becomes
\begin{equation}
{\dot
P_{(0)}}(t)=\frac{(t-a)^{-\alpha}}{\Gamma(1-\alpha)}{}_a\textbf{D}_t^{\alpha}Q(t,x)+{}_a\textbf{D}_t^{\alpha}\frac{dF(Q(t,x))}{dQ(t,x)}.
\end{equation}

 The fractional
Hamiltonian equations are given below

\begin{eqnarray}\label{compact1}
&{\dot
P_{(n)}}(t)&+P_{(n-1)}(t)=\left(\frac{(t-a)^{-\alpha}}{\Gamma(1-\alpha)}Q(t,x)
+\sum_{k=1}^{\infty}\left(
                      \begin{array}{c}
                        \alpha \\
                        k \\
                      \end{array}
                    \right)
                    \frac{(t-a)^{k-\alpha}}{\Gamma(k-\alpha+1)}Q^{(k)}(t,x)\right)\cr
                    &\times&
                    \left(
                      \begin{array}{c}
                        \alpha \\
                        n \\
                      \end{array}
                    \right)
                    \frac{(t-a)^{n-\alpha}}{\Gamma(n-\alpha+1)}\cr
                    &+&
\sum_{k=1}^{\infty}\left(
                      \begin{array}{c}
                        \alpha \\
                        k \\
                      \end{array}
                    \right)
k!\frac{(t-a)^{k-\alpha}}{\Gamma(k-\alpha+1)}\sum_{s=1}^{k}\frac{d\left(F^{(s)}(Q(t,x))\sum\prod_{r=1}^{k}\frac{1}{a_r!}\left(\frac{Q^{(r)}(t,x)}{r!}\right)^{a_r}\right)}{dQ^{(n)}(t,x)},
\end{eqnarray}
and
\begin{equation}\label{chirality}
{\dot Q^{(n)}}(t)=Q^{(n+1)}(t).
\end{equation}
We observe that (\ref{compact1}) can be written in the compact
form as

\begin{eqnarray}
&{\dot
P_{(n)}}(t)&+P_{(n-1)}(t)=\left({}_a\textbf{D}_t^{\alpha}Q(t,x)\right)\left(
                      \begin{array}{c}
                        \alpha \\
                        n \\
                      \end{array}
                    \right)
                    \frac{(t-a)^{n-\alpha}}{\Gamma(n-\alpha+1)}\cr
                    &+&\frac{d ({}_a\textbf{D}_t^{\alpha}F(Q(t,x)))}{d
                    Q^{(n)}(t,x)}, n\in N.
\end{eqnarray}

  We stress on
the fact that x is not related with space [43]. The negative and
the positive values of x corresponds to interactions with the past
and respectively the future [43]. The solution of the chirality
condition (\ref{chirality}), will give $Q(t,x)=q(x+t)$, which
shows the explicit correspondence between q(t) and the left-mover
$Q(x,t)$.

Taking into account that, when $\alpha=1$, $\Gamma(1)=1$,and using
the fact that $\left(\begin{array}{c}
                        1 \\
                        n \\
                      \end{array}\right)=0$, for $n> 1$,$_a\textbf{D}_t^{\alpha}(Q(t,x))=\frac{dQ(x,t)}{dt }$ and the one-to one correspondence between
                      $ q(t)$
                      and the left mover $Q(x,t)$  we conclude  that when the term
${}_a\textbf{D}_t^{\alpha}F(Q(t,x))$ is switched off from (51) and
(54), the classical results are reobtained.

\section{Conclusions}
The fractional variational principles are powerful tools used
successfully in various area of science and engineering. During
the last years several points of view were launched in order to
find the fractional Euler-Lagrange equations and to find an
appropriate fractional Hamiltonian. One of the key point of these
formulations was that under a certain limit the fractional theory
includes the classical. As it is expected the dynamics of the
fractional calculus systems is different from the classical one
but the classical dynamics is recovered as a particular case.

In this paper we investigated the Euler-Lagrange and the
fractional Hamilton equations corresponding to a fractional
generalization of the equivalent Lagrangians. Namely, we add to a
given fractional Lagrangian a term which under certain limit
reproduces the total derivative at the classical level. We have
observed that for the fractional discrete systems the
corresponding generalization of the classical equivalent
Lagrangians lead us to a theory possessing infinite higher order
derivatives. We have calculated the fractional Euler-Lagrangian
for a particular case as well as the fractional Hamilton equations
were obtained in a most general case.

We consider two specific examples in this paper.  For the first
example we consider one of the possible fractional generalization
of the free one dimensional particle Lagrangian and we add to it a
term of the form $C{}_a\textbf{D}_t^{\alpha}q(t)$. From the
classical point of view the above investigated Lagrangians are
equivalent. Both fractional Euler-Lagrange and Hamilton equations
are obtained for this example and it was proved that they are
equivalent in the fractional case. For example, the difference
between the classical case and the fractional one is illustrated
by the form of the fractional Euler-Lagrange equations given by
(27). The second example deals with the same fractional
generalization of the free one dimensional particle but we add a
term of ${}_a\textbf{D}_t^{\alpha}F(q(t))$. In this case, due to
the $Fa\grave{a}$ di Bruno formula, we used the 1+1 field
formalism in order to obtain the corresponding fractional
Euler-Lagrange and Hamilton equations. The classical results are
reobtained when $\alpha=1$.

\section*{Acknowledgments}

{ \small One of the authors (D.B.) would like to thank J. Cresson,
F. Mainardi  for reading the manuscript. The authors would like to
thank J. J. Trujillo for continuing support and for interesting
discussions. This work is partially supported by the Scientific
and Technical Research Council of Turkey.}


\end{document}